# A Generalized Framework for Ontology-Based Information Retrieval

Application to a public-transportation system

Amir ZIDI
L.A.M.I.H. – UMR CNRS 8201
UVHC, Le Mont Houy, 59313 Valenciennes Cedex 9, France.
Amir.zidi@univ-valenciennes.fr

Mourad ABED
L.A.M.I.H. – UMR CNRS 8201
UVHC, Le Mont Houy, 59313 Valenciennes Cedex 9, France.
Mourad.abed@univ-valenciennes.fr

*Abstract*—**In this paper we present a generic framework for ontology-based information retrieval. We focus on the recognition of semantic information extracted from data sources and the mapping of this knowledge into ontology. In order to achieve more scalability, we propose an approach for semantic indexing based on entity retrieval model. In addition, we have used ontology of public transportation domain in order to validate these proposals. Finally, we evaluated our system using ontology mapping and real world data sources. Experiments show that our framework can provide meaningful search results.**

## I. Introduction And Background

The amount of content stored and shared on the Web and other document repositories is increasing fast and continuously. Consequently, the ability to access and select relevant information in these huge and heterogeneous masses of data remains a difficult task. However, most Information retrieval systems have limited abilities to exploit the conceptualizations involved in user needs and content meanings. This involves limitations such as the inability to describe relations between search terms.

In order to overcome these limitations, current Information Retrieval (IR) studies are focusing on relevant documents retrieval using additional knowledge. The main idea is to support a high-level of content and queries conceptual understanding. According to [1], there are two main categories of conceptual-based information retrieval approaches. The first one concerns approaches that extract semantic meaning from documents and queries by analyzing the latent relationships between text words. The second category consists on approaches that, manually or automatically, construct taxonomy of semantic concepts and relations and map documents and queries onto them. Ontology, as a knowledge representation, is one of the most used technologies in the second category. The use of ontology in IR is an important parameter presented by [1] to characterize ontology-based methods. The ontology may be used partially through a query expansion phase [2]. It may also be advanced in both phases of indexing and retrieval. Several approaches exist in the literature such as [3] and [4]. These approaches adopt an advanced use of ontology-based knowledge representation. They can be more efficient especially using domain-information extraction. However, they use specific language for semantic querying which is not easy to be used by the end-users. Formulating a query using such languages requires the knowledge of the domain ontology as well as the syntax of the language.

In this paper, we are focusing on adapting the keyword-based semantic retrieval system using domain ontology in three phases namely the knowledge phase, the indexing phase and the retrieval phase. We are trying to deal with three main issues of the semantic search and retrieval:

- Scalability: it involves not only exploiting semantic metadata that are available in data sources but also managing huge amounts of information having a structured and unstructured content form [5]. In order to achieve more scalability, we propose a semantic indexing approach based on an entity retrieval model.

- Usability: In order to deal with usability issue, we adopt a keyword-based interface as it provides a comfortable and relaxed way to query about the end-user.

- Retrieval performance: we are trying to improve the retrieval performance by using a domain-specific information extraction, inference and rules.

The remainder of this paper is organized as follows: The next section presents the framework of ontology-based information retrieval. This section covers the general architecture and the main processes description including the use of public-transportation ontology, semantic indexing and querying. In order to validate the proposed Framework, the third section includes performed experiments which are based on real word data sources such as RATP open data[1]. The paper concludes with a summary and discussion of the outcomes of the presented work.

## II. PROPOSAL OF FRAMEWORK

Our framework structure is mainly based on three processes: semantic knowledge representation, semantic indexing and semantic querying. The overall diagram of the framework is shown in Fig. 1.We describe the steps we take until the system becomes ready for semantic querying:

- Using the usable information from data source (web sites, data base ...) we populate the initial OWL files.

- We run the Reasoner over these files and obtain new OWL files containing the inferred information.

- We build indexes, using these inferred OWLs, which are used in semantic querying.

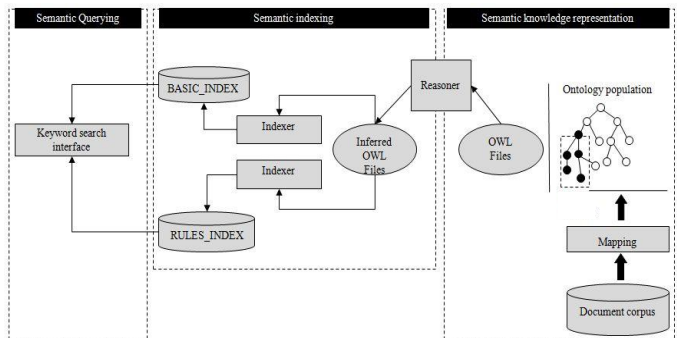

Figure 1.Over all framework diagram

### A. Semantic Knwoledge Representation

Ontology is considered as a key feature to represent semantic knowledge. RDF[2] schema (RDFS[3]), which was built upon RDF, was used to develop ontology language. It extends RDF vocabulary with additional classes and properties such as rdfs:Class and rdfs:subClassOf [3]. OWL[4] further extends RDFS with additional features such as cardinality constraints, equality and disjoint classes, which enable users to better define their classes. In addition to that, OWL classes may be instantiated by adding new individuals. Generally, ontology design is based on the diagram presented in fig.2. This is the diagram of entity types defined for RDF, RDFS and OWL. We can see that user's classes are defined and instantiated based on those entities.

In our work, an existing ontology is reused. It was developed by [6] to facilitate information retrieval for transportation systems. To constitute our knowledge base, we use a wrapper-based method [5]. This latter has as input a data source (data base, www, document corpus). It analyzes and extracts data in order to populate the ontology with instances. The next step is inference. The main idea of this step is to expand knowledge base with new added instances using relations and rules defined by [6]. An example for ontology population can be seen in fig.2, where instances are extracted from RATP open data and Web Annuaire[5]. In this example, we have two new instances (CONNECTION_POINT, OBSERVATOIRE-ASSAS (Paris)) and (SHELTER, Hôtel Istria Montparnasse). After the inference process, we obtain new Knowledge which is OBSERVATOIRE-ASSAS (Paris) is_encercled_by Hôtel Istria Montparnasse. Beyond the relations between classes, authors of the used ontology present a set of rules in order to offer better planning to passengers. As a result, we can have new knowledge about a trip from an origin to a destination.

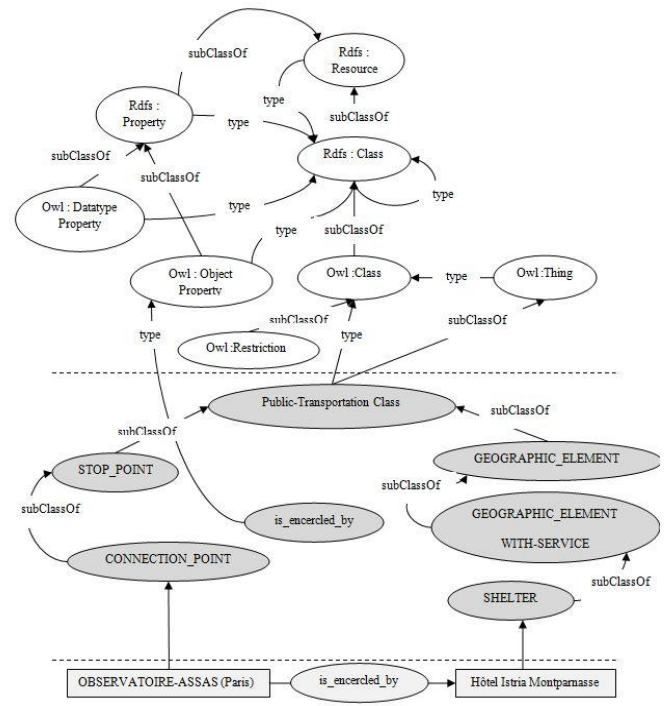

Figure 2.Example of ontology population

After this step, we obtain useful OWL files that will be indexed and used for the search.

### B. Semantic Indexing

As our knowledge base is constituted of entities defined for RDF, RDFs and OWL, we designed an indexing system using entity retrieval model.

#### 1) Entity retrieval model

A knowledge base, which is constituted of entities defined for RDF, is essentially a labeled and directed graph with the nodes being resources while the edges represent the properties [7]. This graph is essentially a set of RDF Triple (N-Triples). An RDF Triple contains three components each of them is providing complementary pieces of information: subject (node), predicate (property) and object (node).

---
[1] http://data.ratp.fr/ http://data.ratp.fr/
[2] http://www.w3.org/RDF/
[3] http://www.w3.org/2001/sw/wiki/RDFS
[4] http://www.w3.org/OWL/
[5] http://www.webannuaire.net/

The subject identifies what object the triple is describing, the predicate defines the piece of data in the object we are giving a value to and the object is the actual value.

In this work, we adopted the Entity Attribute-Value model (EAV model) proposed by [7]. Before describing our indexing system, we estimated useful to first introduce some basic definitions of EAV model. This later is based on a directly labeled graph $G$ which covers the various types of data sources in particular RDF resources.

The graph $G$ represents datasets, entities and their relationships:

$V$: set of nodes

$A$: set of labelled edges

$V^E$: set of entity nodes

$V^L$: set of literal node

$L$: set of labels composed of $L^V$ (set of node labels) and $L^A$ (set of edge labels)

$V^E_D$: set of entity nodes which form a dataset $D$

$L^E_D$: set of entity node labels which form a dataset $D$

$L^V_D$: set of node labels which for a dataset $D$

Graph $G$: is a graph over $L$ and $G=<V, A, \lambda>$ where $\lambda: V \rightarrow L^V$ is node labeling function. The set of labelled edges is defined as $A \subseteq \{(e, \alpha, v)| e \in V^E, \alpha \in L^A, v \in V\}$. The components of edge $a \in A$ is denoted by *source(a)*, *label(a)* and target (a) respectively

A dataset provides information about an entity including its relationships with other entities and its attributes:

Dataset $D$: a dataset over a graph $G=<V, A, \lambda>$ is a tuple $D=<V_D, A_D, L^V_D, \lambda>$ with $V_D \subseteq V$ and $A_D \subseteq A$.

A subgraph describing an entity can be extracted from a dataset; an entity description is defined as:

Tuple $<e, A_e, V_e>$ where $e \in V^E_D$ the entity node, $A_e \subseteq \{(e, \alpha, v)| \alpha \in L^A_D, v \in V_e\}$ the set of labelled edges representing the attributes and $V_e \subseteq V_D$ the set of nodes representing values.

We illustrate an example of an RDF graph extracted from our knowledge base. We can see how dataset are divided into entities description (subgraph).

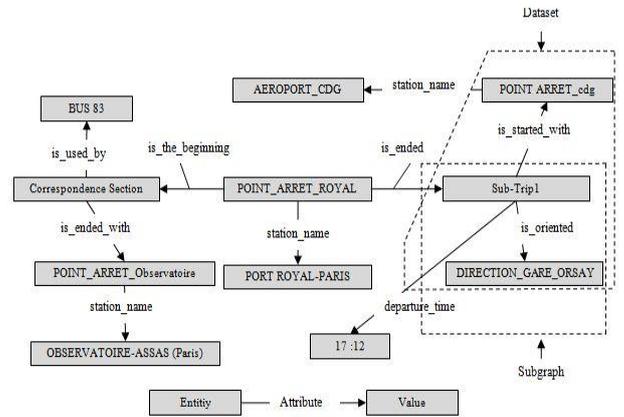

Figure 3. RDF Graph

*2) Index structure*

Retrieval performance depends on the index structure. We constructed two indexes called BASIC_INDEX and RULES_INDEX. The first index (Tab.1) contains all indexed entities which may be retrieved from the knowledge base. While the second contains entities which are inferred using the rules set. As we have mentioned in the previous sections, each entity has its own properties associated with it, such as attribute and value. That information is also included with each entity. Consequently, the structure of each indexed document (e.g. Entity) is composed of four fields <Dataset, Entity, Attribute, Value>. Each field has a name and a text value. While Dataset contains the label of a dataset $D$, Entity contains the label of the entity node $e \in V^E_D$, Attribute label contains the attribute label $\alpha \in L^A_D$ and Value contains the label of the value node. For each RDF triple, Dataset field represents URI set, Entity field represents Subject, Attribute field represents Predicate and Value field represents Object.

| Field | Value |
| --- | --- |
| Dataset | http://www.owlontologies.com/Ontology1256801179.owl#POINT_ARRET_ROYAL |
| Entity | POINT_ARRET_ROYAL |
| Attribute | station_name |
| Value | PORT-ROYAL-Paris |

Table 1. An example of indexing Entity (POINT_ARRET_ROYAL)

We create our second index, RULES_INDEX, which contains all entities generated after the rule inferencing step. In this index, indexed documents are basically a set of journey pattern (Tab.2). This latter is composed of an entity set which may define a trip from origin to a destination. Taking the example of service journey pattern in which banks or post offices are available with the associated connection point. Note that RULES_INDEX is created for retrieval performance purpose.

| Field | Value |
|---|---|
| Dataset | http://www.owlontologies.com/Ontology1256801179.owl#SERVICE_JOURNEY_PATTERN |
| Dataset | http://www.owlontologies.com/Ontology1256801179.owl#POINT_ARRET_observatoire |
| Entity | POINT_ARRET_observatoire |
| Attribute | station_name |
| Value | OBSERVATOIRE_ASSAS (Paris) |
| Attribute | is_ encircled |
| Value | LA_BANQUE_1 |
| Dataset | http://www.owlontologies.com/Ontology1256801179.owl#LA_BANQUE_1 |
| Entity | LA_BANQUE_1 |
| Attribute | nom_element_geographique |
| Value | BANQUE-CENTRALE |

Table 2. An example of indexing a journey pattern (SERVICE_JOURNEY_PATTERN)

## C. Semantic Querying

Once the semantic knowledge is represented and indexed, the next step is querying the EAV graph (e.g. RDF graph). In order to do that, we use SIREn[6], an efficient semi-structured information retrieval for Lucene[7]. Three types of queries are supported:

- Full text: keyword-based query when the data structure is unknown. It allows the user to find all the relevant documents that contain all terms in the query using full-text search syntax.

- Structural: when the data structure is known, it produces precise search results using triple patterns to represent partial or complete triples. A triple pattern is a complete or partial representation of a triple <entity, attribute, value>.

- Semi-structural: combination of the two previous query types when the structure is partially known. Full-text search is supported on any part of the triple, which means that the user can use the Keyword-based query syntax to describe his entity, attribute or value.

### 1) Search with SIREn

With SIREn, Querying RDF graph is commonly achieved using triple stores (i.e. RDF triple, EAV model). We developed a keyword-based interface as it provides a comfortable way to query about the end-user. Query results are achieved using a Boolean combination of attribute-value pairs based on the logical operator ∧, ∨ and ¬, this is called query algebra. In the following we present how we adapted the formal model of relational query algebra, which is used in SIREn and proposed by [7] [8], to our work.

---

[6] http://siren.sindice.com/

[7] http://lucene.apache.org/core/

### 2) Query formulation

In this section, the field Dataset is denoted by $d$, the field Entity is denoted by $e$, the field Attribute is denoted by $at$ and the field Value is denoted by $v$. Given a keyword selection condition $c$ and a relation $R$, the keyword selection operator $\sigma_c(R)$ is defined as a set of relation instances $\{r|r \in R\}$ for which the condition $c$ is true. The condition $c$ consists of testing if a given word denoted by $k$ occurs in one of the field $f$ of a relation $R$, which is denoted by $f{:}k$. we denoted the function of the test by $W$. More details about the function $W$ can be found in [8]. For example if we test if the keyword $k$ occurs in value label of a relation instance r (denoted by $r.v$):

$$\sigma_{v:k}(R): \{r|r \in R, k \in W(r.v)\}$$

We denote by $\pi_f(R)$ the projection operator which allows extracting a specific column of field $f$ from a relation $R$. The projection operator can be used to extract more than one column. For example $\pi_{e,d}(R)$ returns a relation with only two columns, dataset and entity. In the following, we present an example for a simple query formulation, in which, the user is searching for a Hotel Istria.

Q: Find all entities matching keywords *Hotel* and *Istria*.

$Q= \pi_{e,att,v}(\sigma_{v:"Hotel"}(R)) \cap \pi_{e,att,v}(\sigma_{v:"Istria"}(R))$
$Q=\pi_{e,att,v}(\sigma_{v:"Hotel" \wedge v:"Istria"}(R))$

| Entity | Attribute | Value |
|---|---|---|
| OBSERVATOIRE _ASSAS (Paris) | is_encercled_by | Hôtel Istria Montparnasse |

Table 3. An example showing extracted query results using 3 columns

The proofs of used properties can be found in [9].

## III. EVALUATION PROCESS

### A. Evaluation method

In order to evaluate the framework performance, we prepared a set of queries as the example shown in Table.3. We put the corresponding keyword query which was actually used in the evaluation. Then, we calculated the correct number of documents that should be retrieved, for each query. Finally, we run the queries and calculated the performance using evaluation metrics Precision, Recall and the F-Measure. Precision metric is the proportion of the related documents in the retrieved documents (true positives) to the total number of retrieved documents. Recall metric is the proportion of the retrieved related documents to the total number of related documents that should have been retrieved. F-Measure is used as it provides more robust evaluation criteria using Precision and recall together. They are calculated as follows:

$$Precision = \frac{true\ positive}{true\ positive + false\ positive}$$

$$Recall = \frac{\text{true positive}}{\text{true positive} + \text{false negative}}$$

$$F - mesure = 2 * \frac{Precision * Recall}{Precision + Recall}$$

| Q1 | Find the Hotel Istria. (query: "*Hotel Istria*") |
|---|---|
| Q2 | Find a trip from AEROPORT_CDG to PORT-ROYAL-Paris. (query: "*Trip cdg Port-Royal*") |
| Q3 | Find a trip from AEROPORT_CDG to PORT-ROYAL- with a Hotel near to PORT-ROYAL- . (query: "*trip cdg Port-Royal Hotel*") |
| Q4 | Find a trip from AEROPORT_CDG to Hotel Istria (query: "*trip cdg Hotel Istria*") |

Table 4. Example of evaluation queries

Before analyzing the results, we want to clarify the evaluation queries. *Q1* is used to retrieve all entities matching keywords *Hotel* and *Istria*. *Q2* is used to retrieve all entities matching keywords *Trip, cdg* and *Port-Royal*. *Q3* is used to retrieve all entities matching keywords *Trip, cdg* and *Hotel and Port-Royal*. *Q4* is used to retrieve all entities matching keywords *Trip, cdg* and *Hotel Istria*. By executing *Q2*, *Q3* and *Q4*, user should access to all information about a trip from an origin to destination including entities, attributes and values.

*B. Analysis of results*

The obtained results (Tab.5) show that the exploitation of semantic fields shown fruit with high rate of precision and recall. With respect to the precision, scores show that the semantic search presents a high rate. This latter means that little unnecessary documents are provided by our framework and that the latter may be considered as "precise". Additional information (keyword) in *Q3* produced a gain compared to *Q2* and *Q4* in terms of Recall. This gap is also explained by the lack of information about user query. As shown in (Tab.6), this gap can be reduced by separately indexing entities which are generated after the rule inferencing step. Finally, these results are confirmed by the F-measure.

| Queries | Precision (%) | Recall (%) | F-measure |
|---|---|---|---|
| Q1 | 100 | 100 | 1 |
| Q2 | 75,0 | 100 | 0,857 |
| Q3 | 100 | 90,0 | 0,947 |
| Q4 | 100 | 63,0 | 0.777 |

Table 5. Evaluation results (BASIC_INDEX)

.

| Queries | Precision (%) | Recall (%) | F-measure |
|---|---|---|---|
| Q1 | 100 | 88 | 0.936 |
| Q2 | 96,0 | 97,90 | **0,969** |
| Q3 | 95,12 | 98,0 | **0,965** |
| Q4 | 100 | 80,0 | **0.888** |

Table 6. Evaluation results (RULES_INDEX)

IV. CONCLUSION

In this paper, we presented a generic framework for ontology-based information retrieval system and its application in public-transportation domain. We tried to exploit the main advantages of semantic knowledge representation by using a domain-specific information extraction, inference and rules and also to take advantage of semantic indexing to enhance the retrieval performance.

The current implementation can be extended in many ways. We are planning to enrich indexed data by using more meaningful rules to better exploit underlying semantics in content being indexed. In addition, we will focus on a new aspect of a personalized search which integrates user's profile in the indexing phase. The main idea is to re-index contents after clustering user's profiles in order to get more relevant matching between well-defined resources and user queries.